\documentclass[letterpaper, 11pt, journal]{ieeeconf} 

\IEEEoverridecommandlockouts 
\usepackage[utf8]{inputenc}
\usepackage{microtype}
\usepackage{pgfplots}
\usepackage{bm}
\usepackage{mathtools}
\usepackage{enumitem}
\usepackage{breqn}
\usepackage{graphicx,import}
\usepackage[justification=centering]{caption}
\usepackage{subcaption}
\usepackage{float}
\usepackage{soul}  
\usepackage{amssymb}
\usepackage{url}

\pgfplotsset{compat=newest}
\pgfplotsset{plot coordinates/math parser=false}
\pgfplotsset{
/pgfplots/colormap={jet}{rgb255(0cm)=(0,0,128) rgb255(1cm)=(0,0,255)
rgb255(3cm)=(0,255,255) rgb255(5cm)=(255,255,0) rgb255(7cm)=(255,0,0)
rgb255(8cm)=(128,0,0)}
}

\newcommand{\poseBike}{x^b}
\newcommand{\poseRider}{x^r}
\newcommand{\speedBike}{v^b}
\newcommand{\speedRider}{v^r}
\newcommand{\bpath}{\theta^p}

\title{\LARGE \bf The Dynamics of a Bicycle on a Pump Track --\\ First Results on Modeling and Optimal Control
}
\author{Julian Golembiewski, Marcus Schmidt, Benedikt Terschluse, Thomas Jaitner, Thomas Liebig,\\ and Timm Faulwasser*
\thanks{\textbf{Julian Golembiewski}, Institute for Energy Systems, Energy Efficiency and Energy Economics, TU Dortmund University, Dortmund, Germany, e-mail: julian.golembiewski@tu-dortmund.de}%
\thanks{\textbf{Marcus Schmidt}, \textbf{Benedikt Terschluse}, \textbf{Thomas Jaitner}, Institute for Sport and Sport Science, TU Dortmund University, Dortmund, Germany, e-mails: \{marcus2.schmidt, benedikt.terschluse, thomas.jaitner\}@tudortmund.de}%
\thanks{\textbf{Thomas Liebig}, Chair of Artificial Intelligence, TU Dortmund University, Dortmund, Germany, email: thomas.liebig@tudortmund.de}%
\thanks{*\textbf{Corresponding author: Timm Faulwasser}, Institute for Energy Systems, Energy Efficiency and Energy Economics, TU Dortmund University, Dortmund, Germany, e-mail: timm.faulwasser@ieee.org}%
}

\begin{document}
\maketitle
\thispagestyle{empty}
\pagestyle{empty}

\begin{abstract}
   We investigate the dynamics of a bicycle on an uneven mountain bike track split into straight sections with small jumps (kickers) and banked corners. A basic bike-rider model is proposed and used to derive equations of motion, which capture the possibilities to accelerate the bicycle without pedaling. Since this is a first approach to the problem, only corners connected by straight lines are considered to compute optimal riding strategies. The simulation is validated with experimental data obtained on a real pump track. It is demonstrated that the model effectively captures the longitudinal bike acceleration resulting from the relative vertical motion between the rider's upper body and the bicycle. Our numerical results are in good analogy with real rider's actions on similar tracks.
\end{abstract}


\section{Introduction}
\label{cha:intro}
The dynamics of bicycles are challenging due to nonholonomic constraints, soft and hard nonlinearities, and inverse response behavior~\cite{bicycle_teach_system}. Accordingly, bicycle dynamics are a well-known illustrative example for control engineering education~\cite{bicycle_teach_system,Lunze2020}. The lateral dynamics of the rear-wheel steered bicycle, e.g., is frequently drawn upon to showcase non-minimum phase behavior. It is shown in~\cite{bicycle_teach_system,astrom_bicycle_dynamics,wilson_bicycling_science,lunze_fahrrad} that stabilizing such a vehicle is close to impossible for a human rider.

A common choice to analyze bicycle dynamics is the Whipple or Carvallo-Whipple model with its main focus on balancing, stability, and the interplay between speed and stability~\cite{whipple,carvallo}. Many studies have examined or expanded upon the dynamics derived from the Whipple model, particularly with regard to lateral stability~\cite{analysis_carvallo_whipple,papadopoulos_linearize_whipple,xiong_stability_whipple,boyer_reduced_whipple_dynamics,schwab_review_rider_control,phd_human_bicycle_control}.

Here, however, we are not interested in the dynamics of balancing a bicycle rather we consider \textit{the problem of accelerating a bike without pedaling}. At first glance, the ambition to accelerate without pedaling is quite counterintuitive. In cycling, however, it is known that on non-planar surfaces the movement of the rider relative to the bicycle is an efficient means of acceleration, which can be observed, e.g., in Olympic BMX races~\cite{yt_pumpWorldCup,yt_bmxOlympics}.

Despite the considerable research that has been conducted on bicycle modeling, existing studies on the vertical dynamics of bicycle models, including those that concentrate on suspension behavior, system vibrations, and road-tire contact models, such as in~\cite{single_track_dynamics_control,doria_tyres_bicycle,shoman_modeling_bicycle_dynamics}, are not capable of emulating the longitudinal acceleration that a rider can achieve through vertical movement. This reciprocal motion between rider and bicycle is called \textit{pumping} and will serve as our system input throughout this work. This phenomenon can be observed, e.g., on inclines and declines, as well as on steep banked curves. In this work, we focus on the latter only.

The remainder of this paper is structured as follows: A basic bike-rider model with its dynamics in a steep curve is proposed in Section~\ref{cha:modeling}, an optimal control problem formulation for riding through a curve in an optimal manner as well as simulation results are given in Section~\ref{cha:opti}, and Section~\ref{cha:discussion} concludes by discussing the results and validating the suitability of the proposed model. Finally, Section~\ref{cha:conclusion} summarizes the results and gives an outlook on future work.
\section{Modeling}
\label{cha:modeling}
Below, we develop a model of the dynamics of a rider interacting with his bicycle following the initial investigations of~\cite{pumptrack_thesis}. In particular, a basic bike-rider model constrained to a curved surface is proposed. Its dynamics in the form of ordinary differential equations are derived with the Lagrangian formalism according to \cite{fliessbach_lehrbuch_theroetische_physik,bartelmann_theroetische_physik}.  Similar to \cite{particle_surface}, the dynamics are derived using generalized coordinates that are adapted to the surface.

\begin{figure}
    \centering
    \def\svgwidth{175pt}
    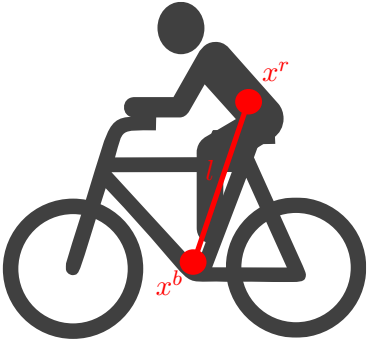
    \caption{Conceptual bike rider model}
    \label{fig:bike_rider_model}
\end{figure}

The fundamental components of our system comprise a \textit{bicycle} and \textit{its rider}. We simplify the models of both by considering each as a point mass and by connecting them using a massless link of length $l$, see Figure~\ref{fig:bike_rider_model}. The second time derivative of the link's length, denoted as $\ddot{l}$, will be utilized as control input for the system at a later stage. We define \begin{equation*}
    \label{eq:bike_pose}
    \poseRider=(x_1^r,x_2^r,x_3^r)^\top
\end{equation*} as the center of mass (CoM) of the rider's body and, analogously, $\poseBike$ as the CoM of the bicycle. Here, $x_1$ corresponds to the $x$-direction while $x_2$ and $x_3$ refer to $y$ and $z$, respectively. \textit{Note}: For the remainder of the paper, the exponent $\star^r$ denotes a parameter belonging to the rider while $\star^b$ denotes a parameter belonging to the bicycle. 

\begin{figure}
    \centering
        \input{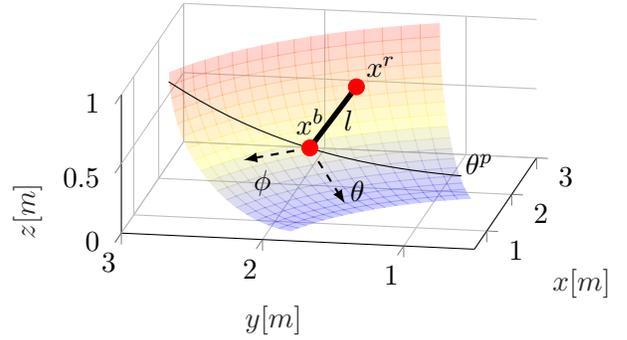}
        \caption{Two-mass model on torus surface}
        \label{fig:bike_in_berm_scetch}
\end{figure}

We model the track as a two-dimensional curved surface $S$ with the parameterization \begin{equation}
    \label{eq:surface_parameterization1}
    g: \mathbb{U}\rightarrow S \text{,}\quad g(\phi,\theta)=\begin{bmatrix}
        (R+r\cos\theta)\cos\phi\\
		(\lambda R+r\cos\theta)\sin\phi\\
		r(1-\sin\theta)
    \end{bmatrix}
\end{equation} where $\mathbb{U}\subset \mathbb{R}^2$ and $S\subset\mathbb{R}^3$. We parametrize an elliptic torus to replicate the shape of a track consisting of two opposing curves connected by two (almost) straight lines. The resulting parameters $\theta$ and $\phi$ therefore define the position of the bicycle in the curve. As it is common for the torus equation, we define $r$ as the radius of the tube, $R$ as the distance from the torus' center to its tube's center, and $\lambda$ as stretching weight. In particular, only a section of the torus’ surface is used and therefore we consider $\theta\in[3/2\pi,2\pi]$. A part of this surface with $\poseBike$ and $\poseRider$ marked is shown in Figure~\ref{fig:bike_in_berm_scetch}. For the sake of simplicity, we consider a specific path \begin{equation}
\label{eq:path}
\bpath(\phi)=\frac{\pi}{2}\cos^2\phi
\end{equation} on the surface. On this path, the cyclist maintains an inner line on the straight sections of the track while shifting towards the outer line at the apex of the curves. This can be observed, for example, in olympic track cycling on a velodrome, where the banked turns allow the cyclists to keep their bikes relatively perpendicular to the surface while riding at high velocities~\cite{wiki_velodrome}. Next, we introduce two simplifications which lead to holonomic constraints for the system: \begin{enumerate}[label=(\alph*)]
	\item \label{ass:dynamics_a} the bicycle's position $\poseBike(t)$ is constrained to the path $\bpath(\phi)$ on the surface $S$ for all $t\in[0,T]$,
	\item \label{ass:dynamics_b} the connection $l(t)$ between the position of the bicycle $\poseBike(t)$ and the rider $\poseRider(t)$ is orthogonal to the surface $S$ for all $t\in[0,T]$.
\end{enumerate}
Consequently, taking item~\ref{ass:dynamics_a} and the surface parameterization \eqref{eq:surface_parameterization1} into account, the position of the bicycle can be expressed by $\phi$: \begin{equation*}
   \label{eq:pose_as_param_bike}
   \poseBike=g(\phi,\bpath)=g(\phi)\text{.}
\end{equation*} For sake of simplicity, item~{\ref{ass:dynamics_b}} restricts the cyclist's position to be upright at all times. This eliminates the degree of freedom to lean forward or backward. Additionally, following item~\ref{ass:dynamics_b} and using elementary geometrical analysis, an analogous expression for the position of the second mass is obtained \begin{equation*}
    \label{eq:pose_as_param_bike}
    \poseRider=\tilde{g}(\phi,l)=\begin{bmatrix}
        (R+r\cos\bpath-l\cos\bpath)\cos\phi \\
		(\lambda R+r\cos\bpath-l\cos\bpath)\sin\phi \\
		r(1-\sin\bpath)+l\sin\bpath
    \end{bmatrix}\text{.}
\end{equation*}
The velocities of $\poseBike$ and $\poseRider$ are given by $\speedBike(t)=\dot{x}^b(t)$ and $\speedRider(t)=\dot{x}^r(t)$, respectively. Furthermore, we set up the Lagrangian in its common form \begin{equation*}
    \label{eq:Lagrangian}
    \mathcal{L}(\phi,\dot{\phi},l,\dot{l})=K(\phi,\dot{\phi},l,\dot{l})-U(\phi,l)\text{,}
\end{equation*} with $K$ denoting the total kinetic energy and $U$ the total potential energy. For two point masses the total kinetic energy is given by the sum of both individual kinetic energies \begin{equation*}
    \label{eq:kinetic_T}
    K=\frac{1}{2}(m^b\|\speedBike\|^2+m^r\|\speedRider\|^2)
\end{equation*} with $m^{\{b,r\}}:=$ mass \{bike, rider\}. Analogously, the total potential energy of two point masses is yielded by the sum of both individual potential energies \begin{equation*}
    \label{eq:potential_U}
    U=g_{\text{grav}}(m^bx_3^b+m^rx_3^r)\text{.}
\end{equation*}
Finally, we use the Euler-Lagrange equations of motion \begin{equation*}
    \label{eq:Euler_Lagrange_equations}
    0 = \frac{d}{dt} \frac{\partial \mathcal{L}}{\partial \dot{\phi}} - \frac{\partial \mathcal{L}}{\partial \phi}\text{,}
\end{equation*} to obtain the dynamics in form of an implicit ordinary differential equation (ODE) \begin{equation}
    \label{eq:system_dynamics}
    0=M(\phi,l)\ddot{\phi} +F(\phi,l){\dot{\phi} }^2 +Q(\phi,l,\dot{l})\dot{\phi}+P(\phi,l,\dot{l},\ddot{l})\text{,}
\end{equation} where $M(\phi,l)$, $F(\phi,l)$, $Q(\phi,l,\dot{l})$, and $P(\phi,l,\dot{l},\ddot{l})$ are given in Appendix A.
\section{Optimal Strategies for Riding a Bicycle through a Pump Track}
\label{cha:opti}
This section introduces the states and inputs of the system under consideration. An optimal control problem (OCP) is formulated to achieve the objective of riding a curve at maximum speed. The constraints of the OCP are derived from real-world experiments. Subsequently, the simulation results are presented.

\subsection{Optimal Control Design}
Consider the state vector \begin{equation*}
  \label{eq:state_vector}
  x=\begin{bmatrix}
    \phi & \dot{\phi} & l &\dot{l}
  \end{bmatrix}^\top\text{,}
\end{equation*}which represents the positions and velocities of the masses relative to the surface, along with the input \begin{equation*}
  \label{eq:input}
  u=\ddot{l}\text{,}
\end{equation*}which characterizes the reciprocal motion between the masses (\textit{pumping}). We obtain the states $\phi$ and $\dot{\phi}$ through the solution of the derived ODE, while $l$ and $\dot{l}$ are acquired by integrating the input $u=\ddot{l}$. The nonlinear dynamics~\eqref{eq:system_dynamics} are an implicit ODE $f(\dot{x}, x, u)=0$. The continuous-time OCP reads\begin{subequations} \label{eq:ocp}
  \begin{align}
    \min_{u(\cdot)\in PC([0,T],\mathbb{R})}& \int_{0}^{T} q^\top x+u^2 \,dt \\
    \text{subject to}\quad\forall t &\in[0,T] \notag \\
    0 &= f(\dot{x}, x, u), \quad x(0) = x_0 \\
    x(t) &\in \mathbb{X} \subseteq \mathbb{R}^4 \\
    u(t) &\in \mathbb{U} \subseteq \mathbb{R}.
  \end{align}
\end{subequations} Here, $x_0$ represents the initial state condition, while the terminal time is denoted by $T$. The vector $q\in\mathbb{R}^{4 \times 1}$ determines the weights assigned to the linear cost term in the states, while the input is penalized quadratically. The weights to penalize $\phi$ and $\dot{\phi}$ are selected as negative. This approach aims to maximize the distance traveled and speed while riding through the curve. Additionally, the closed sets $\mathbb{X}$ and $\mathbb{U}$ represent the constraints on the system states and inputs, respectively.

\subsection{Experimental Results}
\label{sec:exp}
\begin{figure}
  \centering
    \def\svgwidth{0.75\columnwidth}
    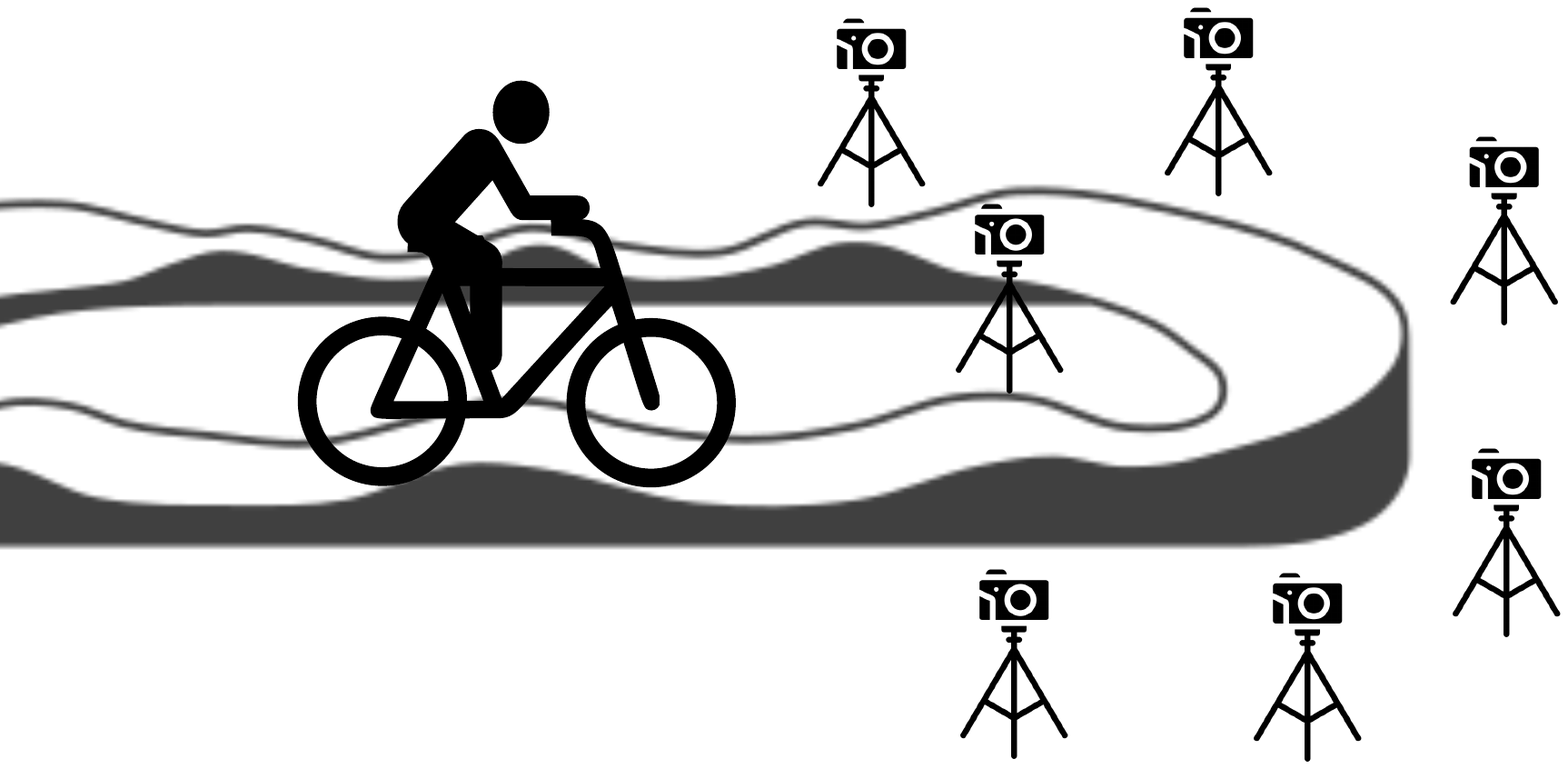
    \caption{Experimental setup}
    \label{fig:experimental_setup}
\end{figure}

\begin{figure}
  \centering
  \includegraphics[width=0.75\columnwidth]{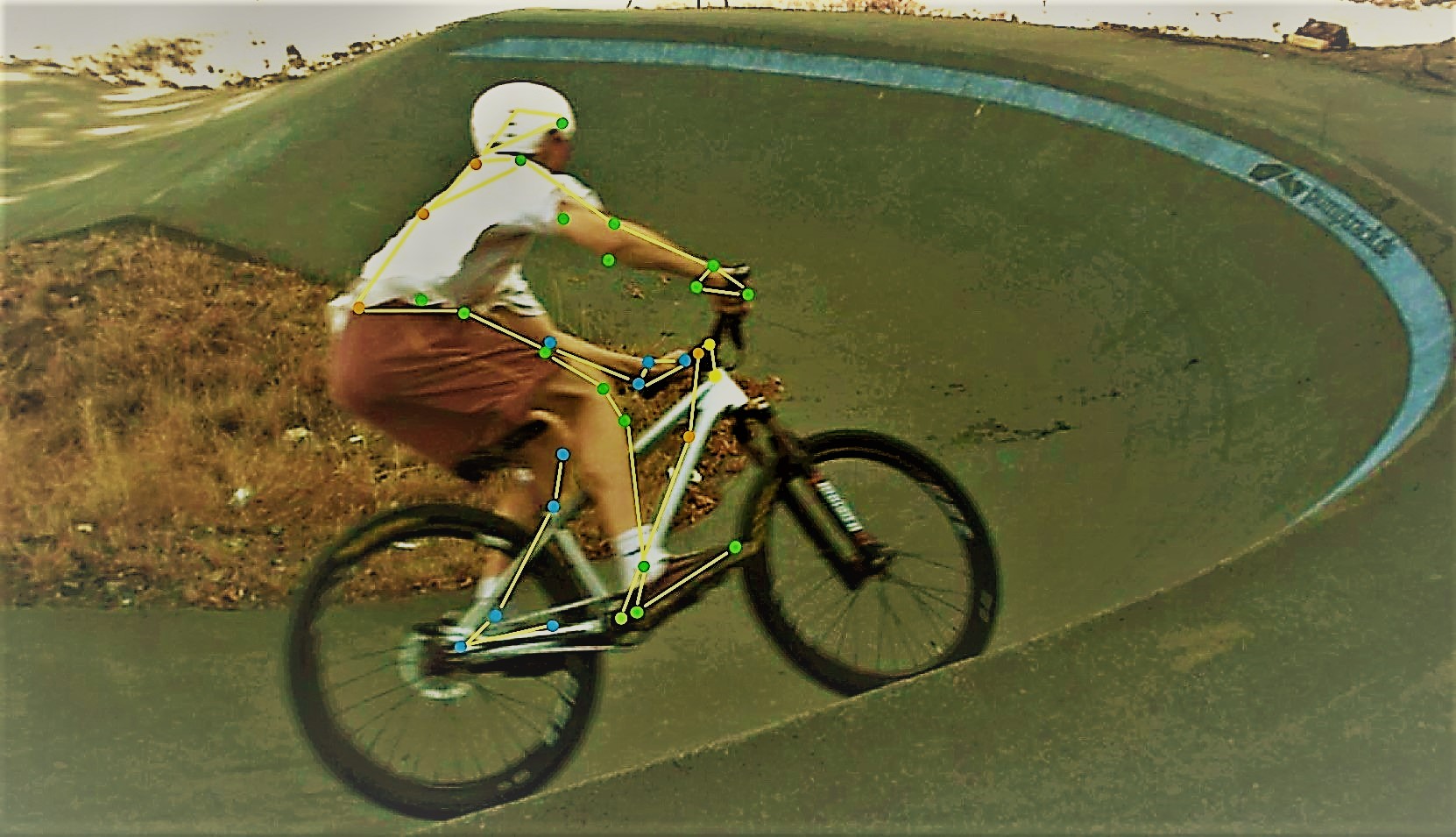}
  \caption{Camera image with marked calculation points}
  \label{fig:marker_sequence}
\end{figure}

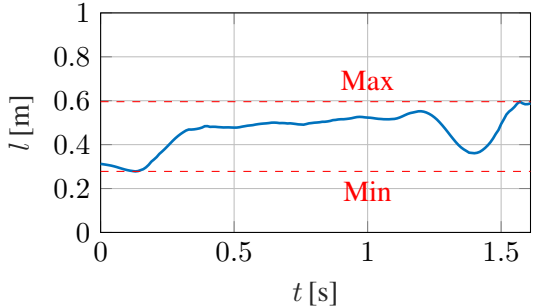
\begin{figure}
  \centering
%
%
\definecolor{mycolor1}{rgb}{0.00000,0.44700,0.74100}%
\begin{tikzpicture}

\begin{axis}[%
width=2.25in,
height=1.15in,
scale only axis,
grid=both,
xmin=0,
xmax=1.61,
xlabel style={font=\color{white!15!black}},
xlabel={$t\,$[s]},
ymin=0,
ymax=1,
ylabel style={font=\color{white!15!black}},
ylabel={$l\,$[m]},
axis background/.style={fill=white},
]
\addplot [color=mycolor1, forget plot, line width=1pt]
  table[row sep=crcr]{%
0	0.312363791644014\\
0.01	0.310483562891347\\
0.02	0.308579731901737\\
0.03	0.306547825877948\\
0.04	0.304237112858426\\
0.05	0.301274035756604\\
0.06	0.297913384142208\\
0.07	0.29410960805662\\
0.08	0.290447063126991\\
0.09	0.286909934139948\\
0.1	0.283867704879412\\
0.11	0.281404567749249\\
0.12	0.27964765008763\\
0.13	0.278028432325324\\
0.14	0.27949923359354\\
0.15	0.283845477321412\\
0.16	0.288209506722963\\
0.17	0.296570960471839\\
0.18	0.307906256762909\\
0.19	0.322535747787189\\
0.2	0.332479749368561\\
0.21	0.343528724115381\\
0.22	0.357811109542951\\
0.23	0.368423202326414\\
0.24	0.379796592558899\\
0.25	0.392058152831353\\
0.26	0.404663042271188\\
0.27	0.417307255608523\\
0.28	0.428955034080047\\
0.29	0.439407715830274\\
0.3	0.44874583803233\\
0.31	0.456859709800339\\
0.32	0.464327705259262\\
0.33	0.468814713850703\\
0.34	0.468589049965413\\
0.35	0.469596808252511\\
0.36	0.471474315267503\\
0.37	0.473997386031648\\
0.38	0.477830701624656\\
0.39	0.483830540632228\\
0.4	0.484184460191894\\
0.41	0.480878001346686\\
0.42	0.481396144111769\\
0.43	0.481166314804027\\
0.44	0.479901793005532\\
0.45	0.479133546069875\\
0.46	0.478214688084235\\
0.47	0.478494162037986\\
0.48	0.478575346396752\\
0.49	0.477956734119475\\
0.5	0.477529610729831\\
0.51	0.477702192068125\\
0.52	0.479253671194983\\
0.53	0.481901597003718\\
0.54	0.483948791809935\\
0.55	0.486178288890271\\
0.56	0.487459411417455\\
0.57	0.488789011972295\\
0.58	0.491369027801985\\
0.59	0.493312865433626\\
0.6	0.494845633994186\\
0.61	0.495966293951249\\
0.62	0.496861921504875\\
0.63	0.498161642759739\\
0.64	0.500004628936433\\
0.65	0.500286657121586\\
0.66	0.499704544130015\\
0.67	0.497976128450851\\
0.68	0.498300514936434\\
0.69	0.497802039071586\\
0.7	0.496361131103559\\
0.71	0.497072485027608\\
0.72	0.495664692127826\\
0.73	0.494121148829259\\
0.74	0.492309439507715\\
0.75	0.491648980717815\\
0.76	0.490833890971144\\
0.77	0.493612048174732\\
0.78	0.496840280916536\\
0.79	0.498231037306356\\
0.8	0.499861509992798\\
0.81	0.500685912564776\\
0.82	0.501874555348171\\
0.83	0.503981899914613\\
0.84	0.504123372887063\\
0.85	0.505660737722867\\
0.86	0.507941709671252\\
0.87	0.511541260479272\\
0.88	0.512539762808947\\
0.89	0.514620270796904\\
0.9	0.515613676800536\\
0.91	0.515616412989798\\
0.92	0.516826613286247\\
0.93	0.518330067769667\\
0.94	0.520269120769541\\
0.95	0.523144625649482\\
0.96	0.525580129453231\\
0.97	0.526447637335521\\
0.98	0.525765381261859\\
0.99	0.52414649107655\\
1	0.523131985330001\\
1.01	0.522478971552711\\
1.02	0.520911597273556\\
1.03	0.519031967051424\\
1.04	0.517508737412315\\
1.05	0.51678548138481\\
1.06	0.515665003225989\\
1.07	0.515470176931399\\
1.08	0.515383208489055\\
1.09	0.516271893773087\\
1.1	0.51698119167778\\
1.11	0.51948149683678\\
1.12	0.52352879189362\\
1.13	0.528714944793489\\
1.14	0.534417812695451\\
1.15	0.53979781743709\\
1.16	0.543852547063772\\
1.17	0.546301957397758\\
1.18	0.549017539383979\\
1.19	0.551723359948693\\
1.2	0.551972381194707\\
1.21	0.550045514916187\\
1.22	0.546189655045288\\
1.23	0.540593861036804\\
1.24	0.533439860769348\\
1.25	0.52434031301857\\
1.26	0.51378689522783\\
1.27	0.501133788959385\\
1.28	0.4875056475053\\
1.29	0.472925007978388\\
1.3	0.457537528546567\\
1.31	0.441751808076394\\
1.32	0.426587092310578\\
1.33	0.411885497757649\\
1.34	0.398633025665618\\
1.35	0.387314288619648\\
1.36	0.378413689687659\\
1.37	0.371060882091133\\
1.38	0.363366717278477\\
1.39	0.362075288512322\\
1.4	0.360831722985605\\
1.41	0.362681674306661\\
1.42	0.367107802853924\\
1.43	0.374953136365793\\
1.44	0.384825349864044\\
1.45	0.396749710999273\\
1.46	0.411052258092265\\
1.47	0.428390761309192\\
1.48	0.448951801878501\\
1.49	0.466499633376976\\
1.5	0.493673366026933\\
1.51	0.514880357000069\\
1.52	0.537359746223665\\
1.53	0.550777372701419\\
1.54	0.560219711724798\\
1.55	0.571841844564875\\
1.56	0.586481773332706\\
1.57	0.595589962783839\\
1.58	0.588237617180851\\
1.59	0.584006402667539\\
1.6	0.58445073579534\\
1.61	0.587815874163007\\
};
\addplot [color=red, dashed, forget plot]
  table[row sep=crcr]{%
0	0.595589962783839\\
1.61	0.595589962783839\\
};
\addplot [color=red, dashed, forget plot]
  table[row sep=crcr]{%
0	0.278028432325324\\
1.61	0.278028432325324\\
};

\node (max) [color=red, above, align=center]
at (axis cs:1, 0.6) {Max};
\node (min) [color=red, below, align=center]
at (axis cs:1, 0.278028432325324) {Min};

\end{axis}
\end{tikzpicture}%
  \caption{Absolute distance between CoM rider and bike down tube}
  \label{fig:exp_results_l}
\end{figure}

\begin{figure}[t]
  \centering
%
%
\definecolor{mycolor1}{rgb}{0.00000,0.44700,0.74100}%
\begin{tikzpicture}
\begin{axis}[%
width=2.25in,
height=1.15in,
scale only axis,
grid=both,
xmin=0,
xmax=1.61,
xlabel style={font=\color{white!15!black}},
xlabel={$t\,$[s]},
ymin=-10,
ymax=35,
ylabel style={font=\color{white!15!black}},
ylabel={$a\,$$[\text{m}/\text{s}^2]$},
axis background/.style={fill=white},
]
\addplot [color=mycolor1, forget plot, line width=1pt]
  table[row sep=crcr]{%
0	7.71226321138884\\
0.01	8.21603646698725\\
0.02	5.69514035766757\\
0.03	2.4964204588525\\
0.04	-0.592329327175957\\
0.05	-2.81881770699316\\
0.06	-4.15369729581502\\
0.07	-4.24174484192527\\
0.08	-3.46800360144526\\
0.09	-2.32143261818634\\
0.1	-1.84736083638687\\
0.11	-2.59754194152421\\
0.12	-4.27554869115042\\
0.13	-6.83904060356925\\
0.14	-7.98214224812335\\
0.15	-7.06269883797672\\
0.16	-4.30858995176514\\
0.17	-1.23699462920768\\
0.18	1.2310578555134\\
0.19	2.59477276269272\\
0.2	3.07360421843147\\
0.21	2.20406819780347\\
0.22	-0.134819296170523\\
0.23	-2.12689152435043\\
0.24	-2.06314939480535\\
0.25	-0.263256201130537\\
0.26	0.995262989599418\\
0.27	0.190815817613335\\
0.28	-1.95548469986882\\
0.29	-4.23328351150684\\
0.3	-5.0527839618841\\
0.31	-4.0548323984057\\
0.32	-1.85456159740159\\
0.33	-0.455812703769041\\
0.34	-0.265446117199144\\
0.35	-0.695551194083452\\
0.36	-0.657199108029511\\
0.37	0.193069121249569\\
0.38	0.45206778574417\\
0.39	0.302151676117291\\
0.4	0.715370820146803\\
0.41	2.02319178963869\\
0.42	3.62245992821711\\
0.43	5.93457359188907\\
0.44	8.37997453517217\\
0.45	10.093388118278\\
0.46	10.7899887170881\\
0.47	10.9847040953641\\
0.48	11.2147357751411\\
0.49	11.704643588683\\
0.5	12.3127208744882\\
0.51	12.8603129858082\\
0.52	12.9174986308069\\
0.53	12.5631462491715\\
0.54	12.3785459297387\\
0.55	12.5069993632684\\
0.56	12.9315069511528\\
0.57	13.6191713321285\\
0.58	14.4967120280744\\
0.59	15.8601226708796\\
0.6	17.6732102451937\\
0.61	19.7426443166464\\
0.62	20.8213222021245\\
0.63	21.0096169194291\\
0.64	21.2416952683034\\
0.65	21.9582695138663\\
0.66	22.4041359756447\\
0.67	22.292615874569\\
0.68	21.9631391088635\\
0.69	21.3714927697178\\
0.7	20.9498114807674\\
0.71	21.0207583722428\\
0.72	21.1378370219066\\
0.73	21.1280947445215\\
0.74	20.7981139136214\\
0.75	19.9257625105801\\
0.76	18.5146929748854\\
0.77	17.198094570413\\
0.78	15.6464616509917\\
0.79	14.4199885207585\\
0.8	14.0401335418234\\
0.81	14.9831266053478\\
0.82	16.6468637185005\\
0.83	18.7600999537447\\
0.84	20.5134877839919\\
0.85	21.8904601076667\\
0.86	21.9492493480165\\
0.87	20.6085445283986\\
0.88	19.1768165020828\\
0.89	18.9519104704768\\
0.9	19.4409723194025\\
0.91	19.1901219830786\\
0.92	17.4815213957342\\
0.93	15.1629128856319\\
0.94	13.3276290582855\\
0.95	12.4542455332441\\
0.96	12.6736453885829\\
0.97	13.3895761573649\\
0.98	14.9474230303153\\
0.99	17.1116317997001\\
1	19.3231366383403\\
1.01	21.2093120973323\\
1.02	23.4188538352031\\
1.03	25.6687730057403\\
1.04	28.1799353660673\\
1.05	29.7475740528346\\
1.06	30.1478116762068\\
1.07	29.7370649874828\\
1.08	29.0180648960414\\
1.09	27.9316863532106\\
1.1	26.3606463627483\\
1.11	24.7995959736743\\
1.12	23.1152353286699\\
1.13	21.4414679315788\\
1.14	19.4816467258895\\
1.15	16.8521356834079\\
1.16	13.2918699273309\\
1.17	9.30321092939881\\
1.18	6.08983945091679\\
1.19	3.74559271495807\\
1.2	2.40135778970568\\
1.21	1.11634688174744\\
1.22	-0.517439143751326\\
1.23	-2.66416481813953\\
1.24	-4.7451194256517\\
1.25	-6.49897524074798\\
1.26	-7.95257744907506\\
1.27	-8.66483516272901\\
1.28	-8.54559119911738\\
1.29	-7.97817752396251\\
1.3	-7.33865480388132\\
1.31	-6.37308923012375\\
1.32	-4.58390549677742\\
1.33	-2.1492667909661\\
1.34	0.185358105150423\\
1.35	1.59933955126727\\
1.36	2.10664314083656\\
1.37	1.79104874416981\\
1.38	0.455564040111945\\
1.39	-1.20209907432131\\
1.4	-2.54328341461146\\
1.41	-3.48431015314646\\
1.42	-3.52200969567119\\
1.43	-2.84887800115398\\
1.44	-1.83787291465097\\
1.45	-0.88419849661662\\
1.46	0.137698671049742\\
1.47	0.869277930267603\\
1.48	1.34489125167461\\
1.49	1.66550574874131\\
1.5	1.83152863145445\\
1.51	1.6269980546007\\
1.52	-0.0157307064279419\\
1.53	-2.80090858709061\\
1.54	-4.08350445046062\\
1.55	-0.490459276281976\\
1.56	7.98427325955303\\
1.57	15.803401241314\\
1.58	16.2809463337761\\
1.59	8.60551182248501\\
1.6	-0.119296240625172\\
1.61	-1.31212785261779\\
};
\addplot [color=red, dashed, forget plot]
  table[row sep=crcr]{%
0	30.1478116762068\\
1.61	30.1478116762068\\
};
\addplot [color=red, dashed, forget plot]
  table[row sep=crcr]{%
0	-8.66483516272901\\
1.61	-8.66483516272901\\
};

\node (max) [color=red, below, align=center]
at (axis cs:1.4, 30) {Max};
\node (min) [color=red, above, align=center]
at (axis cs:0.75, -8.66483516272901) {Min};

\end{axis}
\end{tikzpicture}%
  \caption{Acceleration of the riders CoM relative to the bike's down tube}
  \label{fig:exp_results_a}
\end{figure}
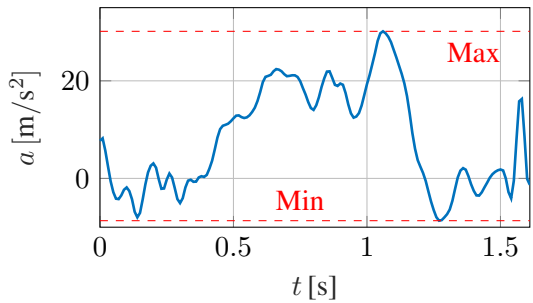

To establish realistic boundaries on the states and inputs of OCP~\eqref{eq:ocp}, real-world experiments were conducted. A ten-camera optoelectrical system (Qualisys\texttrademark) was used to capture a bicycle ride along a steep banked curve at a frame rate of $100\,$Hz, cf. Figure~\ref{fig:experimental_setup}. The rider solely employed pumping motions to generate speed, without any pedaling. Overall, $46$ markers were attached to the main joints of the rider as well as to characteristic landmarks of the bicycle frame, see Figure~\ref{fig:marker_sequence}. The purpose of the experiments was to establish the limitations on the distance $l$ between the CoM of the rider and the bicycle, as well as the system input $u=\ddot{l}$ that denotes the acceleration of that distance. Therefore, the center of mass for the rider was calculated based on a $16$-segment body model considering the relative mass of each segment in relation to the overall body mass. For the bicycle, a fixed marker located at the down tube served as a reference point.
Figures~\ref{fig:exp_results_l} and~\ref{fig:exp_results_a} illustrate the absolute distance and the acceleration between the CoM of the rider and the bicycle's down tube, respectively. The measurements capture a single curve ride.
The horizontal lines highlight the maximum and minimum values of significance and lead to the following constraints \begin{subequations}
  \label{eq:exp_constraints}
  \begin{align}
    0.27803\,\text{m}\, \leq\,\, &l \leq\, 0.59559\,\text{m} \\
    -8.6648\,\text{m}/\text{s}^2\, \leq\,\, &\ddot{l} \leq\, 30.1478\,\text{m}/\text{s}^2\text{.}
  \end{align}
\end{subequations}

\subsection{Simulation Results}
\begin{figure*}
  \centering
  \input{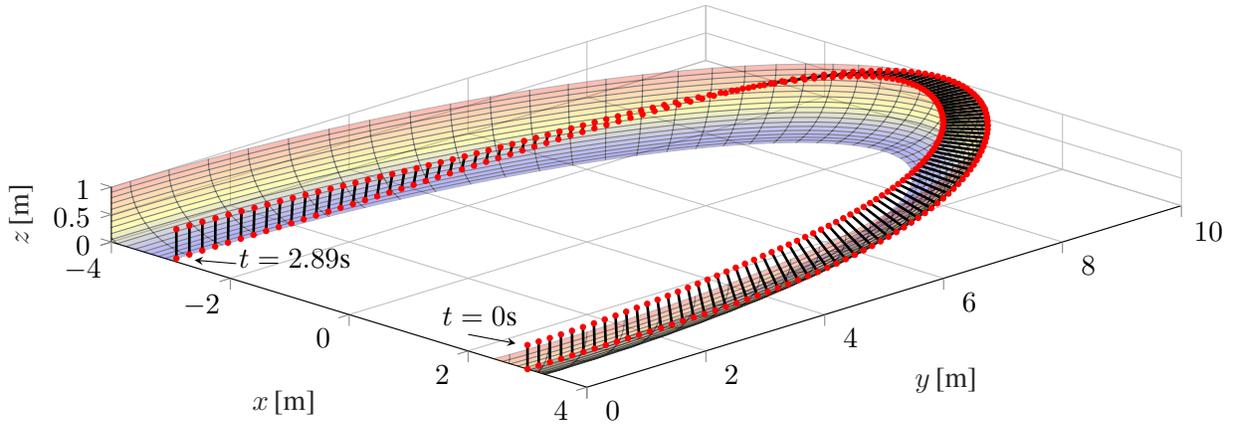}
  \caption{Visualization of position within first curve}
  \label{fig:states_visualized}
\end{figure*}

The following section presents the numerical results obtained from solving OCP~\eqref{eq:ocp} including the constraints~\eqref{eq:exp_constraints}. For the objective function, we select $q=\begin{bmatrix}
  -65 & -65 & 0 & 0 \\
  \end{bmatrix}^\top$. The weights for the states $\phi$ and $\dot{\phi}$ are selected to be negative. Since $\phi$ denotes the evolution of the bike-rider model through the curve, the optimization problem therefore aims to identify the optimal system input $u^\star$ that maximizes the speed and distance traveled along this curve. We solve the OCP with $T=5\,$s. The implicit dynamics are reformulated to an explicit form and discretized with the Runge-Kutta method (RK4) and an integration step size $h=0.01\,$s. The initial state of the dynamics is given by $x_0=[0, \pi/3, (l_{max}+l_{min}/2), 0]$, where $\phi(0)=0$ indicates that the bicycle starts at the midpoint between the opposing curves, and $\dot{\phi}(0)=\pi/3\,$rad/s corresponds to an initial bicycle speed of $9.43\,$m/s. Figure~\ref{fig:states_visualized} depicts a visual representation of the system's pose at time $t=0$. The constraints on the state $l$ and the input $\ddot{l}$ are set according to the values in~\eqref{eq:exp_constraints}, while the remaining states are unconstrained. The nonlinear program is set up in \textit{MATLAB}~\cite{matlab} via \textit{CasADi}~\cite{casad} and solved using \textit{IPOPT}~\cite{ipopt}. The physical parameters of the model are presented in Table~\ref{tab:params}. \begin{table}
  \begin{center}
  \begin{tabular}{c|c|c|c|c|c} 
   $m^b$ & $m^r$ & $g_{grav}$ & $R$ & $r$ & $\lambda$\\ [0.5ex] 
   \hline
   $15\,$\normalfont kg & $80\,$\normalfont kg & $9.8067\,$\normalfont m/s & $3\,$\normalfont m & $1\,$\normalfont m & $3$\\ [0.5ex] 
  \end{tabular}
  \caption{Physical model parameters}
  \label{tab:params}
  \end{center}
  \vspace{-0.8cm}
\end{table}
  
\begin{figure*}[t]
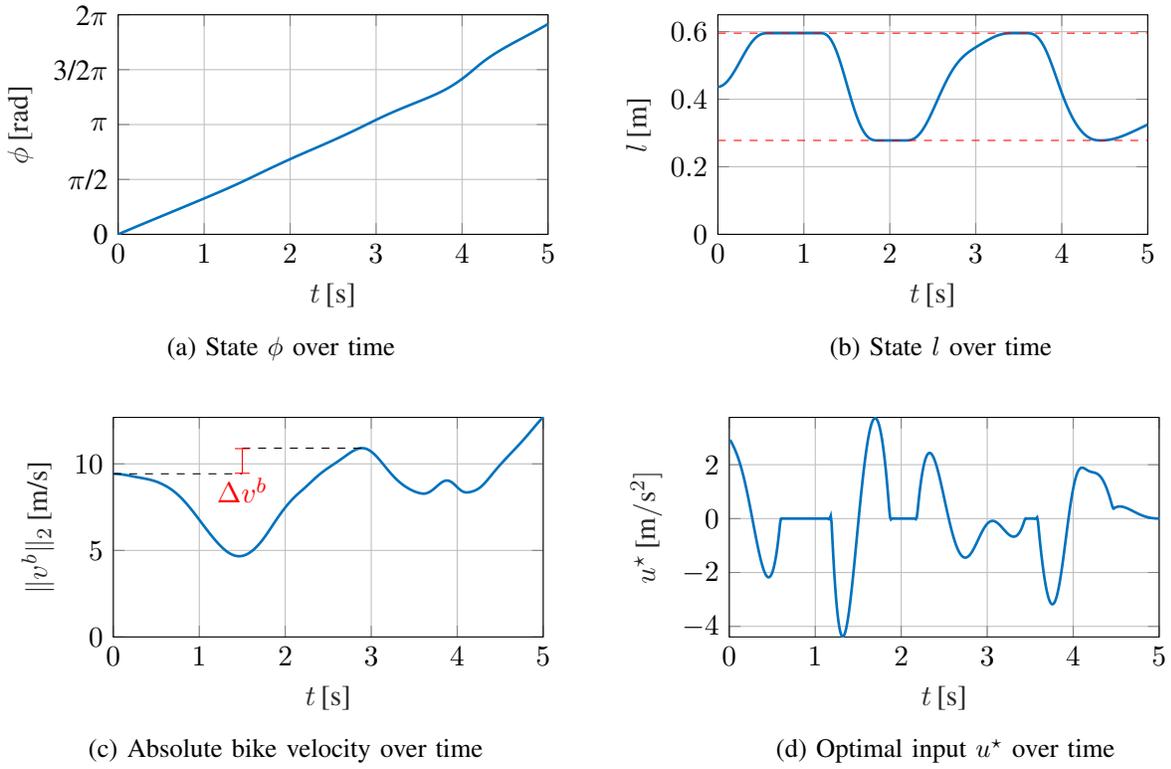

  \centering
 \begin{subfigure}{\columnwidth}
  \centering
  \input{Graphics/tikz/figure_phi_time.tex}  
  \caption{State $\phi$ over time}
  \label{fig:phi-t}
\end{subfigure} \begin{subfigure}{\columnwidth}
  \input{Graphics/tikz/figure_l_time.tex}  
  \caption{State $l$ over time}
  \label{fig:l-t}
\end{subfigure}
\newline
\begin{subfigure}{\columnwidth}
  \vspace{0.7cm}
  \centering
  \input{Graphics/tikz/figure_vel_time.tex}   
  \caption{Absolute bike velocity over time}
  \label{fig:v-t}
\end{subfigure}
\begin{subfigure}{\columnwidth}
  \vspace{0.7cm}
  \input{Graphics/tikz/figure_ddl_time.tex}  
  \caption{Optimal input $u^\star$ over time}
  \label{fig:ddl-t}
 \end{subfigure}
 \caption{Simulation results}
 \label{fig:sim_results}
\end{figure*}

 The simulation results are presented in Figure~\ref{fig:sim_results}. Figure~\ref{fig:phi-t} demonstrates that $\phi(T)$ approaches approximately $2\pi$, indicating that a complete cycle through the track is simulated, successfully covering both steep curves. Figure~\ref{fig:v-t} displays the magnitude of the bicycle velocity over time. The speed difference, $\Delta \speedBike$, before and after the first curve is highlighted. This comparison is made at the straight sections of the track at $t=0\,$s and $t=2.89\,$s, cf. Figure \ref{fig:states_visualized}. Remarkably, we observe a total speed increase of $\Delta \speedBike=1.49\,$m/s, solely generated by the reciprocal motion between the masses.

The input for this reciprocal motion is depicted in Figure~\ref{fig:ddl-t}, while the distance between the masses is illustrated in Figure~\ref{fig:l-t}. Here, we observe that the state $l$ alternates between its constraint values $l_{\text{max}}$ and $l_{\text{min}}$ at specific points along the track. By referring to the visualization in Figure~\ref{fig:states_visualized}, it becomes evident that in order to generate speed, the bike-rider system enters the curve at its maximum height and exits at its minimum height. This behavior is consistent with the second curve as well, where another increase in speed can be observed.

To further support this effect, we conduct an additional comparison by simulating the time required to reach the terminal state $\phi(T)$ without providing any input to the system. It is important to note that the simulations with input $u=0$ are performed with varying initial states for the distance $l$. Interestingly, it is observed that simulating with the maximum value $l_{\text{max}}$ results in the fastest motion. In this case, the system requires a total time of $t=6.13\,$s to reach the same distance. In contrast, when applying the optimal pumping input $u^\star$ to the system, the time taken to complete a full lap through both curves is reduced to $5\,$s. This reveals a significant difference of $\Delta t=1.13\,$s between the two simulations. In other words, by applying the optimal input $u^\star$ to the system, we achieve an $18.43\,$\% reduction in the time required to complete a full lap through the curves.
\section{Discussion}
\label{cha:discussion}
The proposed model successfully captures the relationship between body positioning and longitudinal acceleration during curve riding. It is shown that adopting an extended upper body position when entering a curve, cf. Figure~{\ref{fig:l-t}} at $t\approx1.2\,$s, and transitioning to a shortened position throughout the curve, cf. Figure~{\ref{fig:l-t}} at $t\approx1.5s$, results in longitudinal acceleration. In fact, the pumping motion decreased the time needed for riding a full lap in the simulation by a significant amount. This observation aligns with the real-world experiment conducted in Section~\ref{sec:exp}, where a similar movement pattern was observed for the test rider. Figure~\ref{fig:exp_results_l} clearly illustrates the rise in $l$ as the rider enters the curve, followed by a subsequent decrease upon exiting it. The obtained results are therefore consistent with empirical observations of real athletes to a certain extent. However, it is important to acknowledge that the simplicity of the two-mass model, along with the simplifications imposed on the system, leads to the exclusion of various realistic aspects. These aspects include the intricate physics of the human body, the reduction to a single ground contact point, considerations of ground contact physics, and limitations imposed by the given path~\eqref{eq:path} and the rider's permanent upright position.

Although incorporating these phenomena would enhance the agreement between the model and experiment, the proposed model adequately highlights a gap in the research on bicycle modeling. By extending the analysis to include hilly terrain, where riders have the opportunity to jump on inclines and transition smoothly into declines, the influence of pumping motions on speed generation becomes even more significant.

\section{Conclusion and Outlook}
\label{cha:conclusion}
This paper has proposed a fundamental bike-rider model for analyzing bicycle dynamics on uneven mountain bike tracks in the absence of pedaling. The model's dynamics are derived and utilized to determine the optimal input strategies for riding steep banked curves. Specifically, an OCP was designed to maximize the velocity and distance traveled throughout the curve. The simulation results demonstrate a notable velocity gain, highlighting the relationship between the rider's reciprocal motion and longitudinal acceleration, consistent with real-world athlete observations.

Further exploration of the proposed model on hilly terrain presents an avenue for future research. The dynamics of the bicycle-rider system on uneven tracks become non-smooth due to the presence of jumps, which introduce a distinction between air and ground dynamics~\cite{pumptrack_thesis}. Consequently, addressing this scenario would entail solving OCPs for a system with state jumps to generate optimal riding strategies, thereby introducing an intriguing analytical challenge.

Finally, additional analyses of neuromuscular motor control strategies and intrinsic dynamics of the rider's movement would provide deeper insight into if and how theoretically derived optimal riding strategies could be realized by the human rider. This would further allow us to address practically highly relevant aspects of performance improvement as well as of mechanical load management with regard to injury prevention.


\section*{Acknowledgement}
We would like to give a special thanks to \textit{Jan Dieckmann}, who supported this work as a test rider for all of our experiments. Furthermore, we would like to thank \textit{Dominik Modenbach}, \textit{Gerrit Lang}, and \textit{Maximilian Sueck} for their support in data collection and analysis.


\onecolumn
\appendix
\section{A. Equations of Motion}
A. Equations of Motion:\\

\label{appendix_a}
\begin{minipage}{\textwidth}
    Terms for the equations of motions \eqref{eq:system_dynamics}. We use the short hands $s(x):=\sin(x)$ and $c(x):=\cos(x)$:
    \begin{equation*}
        \begin{split}
            M(\phi,l)&=\frac{m^b}{2} {\left({\sigma_{10} }^2 2+2{\sigma_{11} }^2 +2r^2 \pi^2 {c \left(\sigma_{25} \right)}^2 {c \left(\phi \right)}^2 {s \left(\phi \right)}^2 \right)}+\frac{m^r}{2} {\left(2{\sigma_1 }^2 +2{\sigma_2 }^2 +2{\sigma_3 }^2 \right)}\\
            F(\phi,l)&=\frac{1}{2}\left(-m^b \left(2\sigma_{11} \sigma_9 -\sigma_{10} \sigma_8 2-2r^2 \pi^2 {c \left(\sigma_{25} \right)}^2 c \left(\phi \right){s \left(\phi \right)}^3+2r^2 \pi^2 {c \left(\sigma_{25} \right)}^2 {c \left(\phi \right)}^3 s \left(\phi \right)\right. \right.\\
            &\quad\quad\quad \left. +2r^2 \pi^3 c \left(\sigma_{25} \right)s \left(\sigma_{25} \right){c \left(\phi \right)}^3 {s \left(\phi \right)}^3 \right)+m^b \left(4\sigma_{11} \sigma_9 -\sigma_{10} \sigma_8 4-4r^2 \pi^2 {c \left(\sigma_{25} \right)}^2 c \left(\phi \right){s \left(\phi \right)}^3 \right. \\
            &\quad\quad\quad \left. +4r^2 \pi^2 {c \left(\sigma_{25} \right)}^2 {c \left(\phi \right)}^3 s \left(\phi \right)+4r^2 \pi^3 c \left(\sigma_{25} \right)s \left(\sigma_{25} \right){c \left(\phi \right)}^3 {s \left(\phi \right)}^3 \right)\\
            &\quad\quad\quad \left.+m^r {\left(\sigma_1 \sigma_4 2-2\sigma_3 \sigma_{23} +\sigma_2 \sigma_5 2\right)}-m^r {\left(-2\sigma_2 \sigma_7 +\sigma_1 \sigma_4 2-4\sigma_3 \sigma_{23} +\sigma_2 \sigma_5 2+2\sigma_1 \sigma_6 \right)}\right)\\
            Q(\phi,l,\dot{l})&=\frac{1}{2}\left(-m^r \left(\sigma_1 \sigma_{21} 2+\sigma_1 \left(2\dot{l} \pi s \left(\sigma_{25} \right)c \left(\phi \right){s \left(\phi \right)}^2 +2\dot{l} c \left(\sigma_{25} \right)c \left(\phi \right)\right)2+2\sigma_2 \sigma_{22} \right. \right.\\
            &\quad\quad\quad +2\sigma_2 {\left(2\dot{l} c \left(\sigma_{25} \right)s \left(\phi \right)-2\pi \dot{l} s \left(\sigma_{25} \right){c \left(\phi \right)}^2 s \left(\phi \right)\right)}+\sigma_{15} -2\dot{l} c \left(\sigma_{25} \right)s \left(\phi \right)\sigma_6 -2\dot{l} c \left(\sigma_{25} \right)c \left(\phi \right)\sigma_7\\
            &\quad\quad\quad \left. -6\pi \dot{l} c \left(\sigma_{25} \right)c \left(\phi \right)s \left(\phi \right)\sigma_3 \right)+m^r \left(\sigma_1 \sigma_{21} 2+2\sigma_2 \sigma_{22} +\sigma_{15} -\dot{l} c \left(\sigma_{25} \right)s \left(\phi \right)\sigma_4 2+\dot{l} c \left(\sigma_{25} \right)c \left(\phi \right)\sigma_5 2\right.\\
            &\quad\quad\quad \left. \left.-2\pi \dot{l} c \left(\sigma_{25} \right)c \left(\phi \right)s \left(\phi \right)\sigma_3 \right)\right)\\
            P(\phi,l,\dot{l},\ddot{l})&=-g_{\text{grav}}\left(m^r \sigma_3 -\pi m^b rc \left(\sigma_{25} \right)c \left(\phi \right)s \left(\phi \right)\right)-\frac{m^r}{2} \left(2\ddot{l} s \left(\sigma_{25} \right)\sigma_3 -2\ddot{l} c \left(\sigma_{25} \right)c \left(\phi \right)\sigma_2 -\sigma_{12} +\sigma_{13}\right.\\
            &\quad\quad\quad \left.+2\ddot{l} c \left(\sigma_{25} \right)s \left(\phi \right)\sigma_1 +\sigma_{14} \right)+\frac{m^r}{2}\left(-\sigma_{12} +\sigma_{13} +\sigma_{14} \right)
        \end{split}
    \end{equation*}
    \setlength\jot{0pt}
    \begin{minipage}{.49\textwidth}
        with
        \begin{equation*}
            \begin{split}
                \sigma_1 &=c \left(\phi \right){\left(R\lambda -\sigma_{24} \right)}-\pi s \left(\sigma_{25} \right)c \left(\phi \right){s \left(\phi \right)}^2 {\left(l-r\right)}\\
                \sigma_2 &=s \left(\phi \right){\left(R-\sigma_{24} \right)}+\pi s \left(\sigma_{25} \right){c \left(\phi \right)}^2 s \left(\phi \right){\left(l-r\right)}\\
                \sigma_3 &=\pi lc \left(\sigma_{25} \right)c \left(\phi \right)s \left(\phi \right)-\pi rc \left(\sigma_{25} \right)c \left(\phi \right)s \left(\phi \right)\\
                \sigma_4 &=\sigma_{17} -s \left(\phi \right)\sigma_{16} +2\pi s \left(\sigma_{25} \right){c \left(\phi \right)}^2 s \left(\phi \right){\left(l-r\right)}\\
                \sigma_5 &=-\sigma_{18} +c \left(\phi \right)\sigma_{16} +2\pi s \left(\sigma_{25} \right)c \left(\phi \right){s \left(\phi \right)}^2 {\left(l-r\right)}\\
                \sigma_6 &=\sigma_{17} -\pi s \left(\sigma_{25} \right){s \left(\phi \right)}^3 {\left(l-r\right)}\\
                &-\pi^2 c \left(\sigma_{25} \right){c \left(\phi \right)}^2 {s \left(\phi \right)}^3 {\left(l-r\right)}\\
                &+3\pi s \left(\sigma_{25} \right){c \left(\phi \right)}^2 s \left(\phi \right){\left(l-r\right)}\\
                \sigma_7 &=\sigma_{18} +\pi s \left(\sigma_{25} \right){c \left(\phi \right)}^3 {\left(l-r\right)}\\
                &-\pi^2 c \left(\sigma_{25} \right){c \left(\phi \right)}^3 {s \left(\phi \right)}^2 {\left(l-r\right)}\\
                &-3\pi s \left(\sigma_{25} \right)c \left(\phi \right){s \left(\phi \right)}^2 {\left(l-r\right)}\\
                \sigma_8 &=s \left(\phi \right)\sigma_{19} +\pi rs \left(\sigma_{25} \right){s \left(\phi \right)}^3\\
                &+r\pi^2 c \left(\sigma_{25} \right){c \left(\phi \right)}^2 {s \left(\phi \right)}^3-3\pi rs \left(\sigma_{25} \right){c \left(\phi \right)}^2 s \left(\phi \right)\\
                \sigma_9 &=c \left(\phi \right)\sigma_{20} -\pi rs \left(\sigma_{25} \right){c \left(\phi \right)}^3\\
                &+r\pi^2 c \left(\sigma_{25} \right){c \left(\phi \right)}^3 {s \left(\phi \right)}^2+3\pi rs \left(\sigma_{25} \right)c \left(\phi \right){s \left(\phi \right)}^2 \\
                \sigma_{10} &=\pi rs \left(\sigma_{25} \right)c \left(\phi \right){s \left(\phi \right)}^2 +c \left(\phi \right)\sigma_{19} \\
                \sigma_{11} &=s \left(\phi \right)\sigma_{20} -\pi rs \left(\sigma_{25} \right){c \left(\phi \right)}^2 s \left(\phi \right)
            \end{split}
        \end{equation*}
    \end{minipage}\hfill
    \begin{minipage}{.49\textwidth}
        \begin{equation*}
            \begin{split}
                \sigma_{12} &=2\dot{l} c \left(\sigma_{25} \right)s \left(\phi \right)\sigma_{21}\\
                \sigma_{13} &=2\dot{l} c \left(\sigma_{25} \right)c \left(\phi \right)\sigma_{22} \\
                \sigma_{14} &=2\pi {\dot{l} }^2 c \left(\sigma_{25} \right)s \left(\sigma_{25} \right)c \left(\phi \right)s \left(\phi \right)\\
                \sigma_{15} &=2\dot{l} s \left(\sigma_{25} \right)\sigma_{23} \\
                \sigma_{16} &=\pi s \left(\sigma_{25} \right){s \left(\phi \right)}^2 {\left(l-r\right)}-\pi s \left(\sigma_{25} \right){c \left(\phi \right)}^2 {\left(l-r\right)}\\
                &+\pi^2 c \left(\sigma_{25} \right){c \left(\phi \right)}^2 {s \left(\phi \right)}^2 {\left(l-r\right)}\\
                \sigma_{17} &=s \left(\phi \right){\left(R\lambda -\sigma_{24} \right)}\\
                \sigma_{18} &=c \left(\phi \right){\left(R-\sigma_{24} \right)}\\
                \sigma_{19} &=rc \left(\sigma_{25} \right)+R\lambda \\
                \sigma_{20} &=R+rc \left(\sigma_{25} \right)\\
                \sigma_{21} &=\pi \dot{l} s \left(\sigma_{25} \right)c \left(\phi \right){s \left(\phi \right)}^2 +\dot{l} c \left(\sigma_{25} \right)c \left(\phi \right)\\
                \sigma_{22} &=\dot{l} c \left(\sigma_{25} \right)s \left(\phi \right)-\pi \dot{l} s \left(\sigma_{25} \right){c \left(\phi \right)}^2 s \left(\phi \right)\\
                \sigma_{23} &=\pi lc \left(\sigma_{25} \right){c \left(\phi \right)}^2 -\pi rc \left(\sigma_{25} \right){c \left(\phi \right)}^2\\
                &-\pi lc \left(\sigma_{25} \right){s \left(\phi \right)}^2+\pi rc \left(\sigma_{25} \right){s \left(\phi \right)}^2\\
                &+l\pi^2 s \left(\sigma_{25} \right){c \left(\phi \right)}^2 {s \left(\phi \right)}^2\\
                &-r\pi^2 s \left(\sigma_{25} \right){c \left(\phi \right)}^2 {s \left(\phi \right)}^2\\
                \sigma_{24} &=c \left(\sigma_{25} \right){\left(l-r\right)}\\
                \sigma_{25} &=\frac{\pi {c \left(\phi \right)}^2 }{2}
            \end{split}
        \end{equation*}
    \end{minipage}
\end{minipage}
\twocolumn


\bibliographystyle{unsrt}
\bibliography{Literature/Literatur}

\begin{thebibliography}{10}

\bibitem{bicycle_teach_system}
R.~Klein.
\newblock Using bicycles to teach system dynamics.
\newblock {\em IEEE Control Systems Magazine}, 9(3):4--9, 4 1989.

\bibitem{Lunze2020}
J.~Lunze.
\newblock {\em Regelungstechnik 1}.
\newblock Springer Berlin Heidelberg, 2020.

\bibitem{astrom_bicycle_dynamics}
K.~J. Astrom, R.~E. Klein, and A.~Lennartsson.
\newblock Bicycle dynamics and control: adapted bicycles for education and
  research.
\newblock {\em {IEEE} Control Systems}, 25(4):26--47, aug 2005.

\bibitem{wilson_bicycling_science}
D.~Wilson.
\newblock {\em Bicycling science}.
\newblock MIT Press, Cambridge, Mass, 2004.

\bibitem{lunze_fahrrad}
K.~J. Astr{\"o}m and J.~Lunze.
\newblock Warum k{\"o}nnen wir fahrrad fahren?
\newblock {\em at Automatisierungstechnik 49}, 2001.

\bibitem{whipple}
F.~J.~W. Whipple.
\newblock The stability of the motion of a bicycle.
\newblock {\em Quarterly Journal of Pure and Applied Mathematics},
  30(120):312--348, 1899.

\bibitem{carvallo}
E.~Carvallo.
\newblock Th{\'e}orie du movement du monocycle, part 2: Th{\'e}orie de la
  bicyclette.
\newblock {\em J. Ec. Polytech. Paris}, 6:1--118, 1901.

\bibitem{analysis_carvallo_whipple}
T.~Tun, L.~Rothenbusch, P.~Ingenlath, A.~Brezing, and B.~Corves.
\newblock Modelling, implementation and analysis of the carvallo-whipple
  bicycle model in msc adams.
\newblock 2018.

\bibitem{papadopoulos_linearize_whipple}
J.~P. Meijaard, J.~M. Papadopoulos, A.~Ruina, and A.~L. Schwab.
\newblock Linearized dynamics equations for the balance and steer of a bicycle:
  a benchmark and review.
\newblock {\em Proceedings of the Royal society A: mathematical, physical and
  engineering sciences}, 463(2084):1955--1982, 2007.

\bibitem{xiong_stability_whipple}
J.~Xiong, N.~Wang, and C.~Liu.
\newblock Stability analysis for the whipple bicycle dynamics.
\newblock 48(3):311--335, oct 2019.

\bibitem{boyer_reduced_whipple_dynamics}
F.~Boyer, M.~Porez, and J.~Mauny.
\newblock Reduced dynamics of the non-holonomic whipple bicycle.
\newblock {\em Journal of Nonlinear Science}, 28(3):943--983, 2018.

\bibitem{schwab_review_rider_control}
A.~L. Schwab and J.~P. Meijaard.
\newblock A review on bicycle dynamics and rider control.
\newblock 51(7):1059--1090, jul 2013.

\bibitem{phd_human_bicycle_control}
J.~K. Moore.
\newblock {\em Human control of a bicycle}.
\newblock PhD thesis, University of California, Davis, 2012.

\bibitem{yt_pumpWorldCup}
Velosolutions.
\newblock Red {B}ull {UCI} {P}ump {T}rack {W}orld {C}hampionships {W}orld
  {F}inal {L}ivestream [{V}ideo].
\newblock https://www.youtube.com/watch?v=f-YgI50wWvQ, 2022.
\newblock Acessed: 17.05.2023.

\bibitem{yt_bmxOlympics}
Olympics.
\newblock Men's {BMX} {G}old {M}edal {R}ace {T}okyo {R}eplays [{V}ideo].
\newblock https://www.youtube.com/watch?v=2RlLmK5WM_s, 2022.
\newblock Acessed: 17.05.2023.

\bibitem{single_track_dynamics_control}
M.~Corno, G.~Panzani, and S.~M. Savaresi.
\newblock Single-track vehicle dynamics control: State of the art and
  perspective.
\newblock {\em IEEE/ASME Transactions on Mechatronics}, 20(4):1521--1532, 2015.

\bibitem{doria_tyres_bicycle}
A.~Doria, M.~Tognazzo, G~Cusimano, V.~Bulsink, A.~Cooke, and B.~Koopman.
\newblock Identification of the mechanical properties of bicycle tyres for
  modelling of bicycle dynamics.
\newblock 51(3):405--420, mar 2013.

\bibitem{shoman_modeling_bicycle_dynamics}
M.~Shoman and H.~Imine.
\newblock Modeling and simulation of bicycle dynamics.
\newblock In {\em Proc. TRA}, pages 1--10, 2020.

\bibitem{pumptrack_thesis}
J.~Golembiewski.
\newblock Modeling and analysis of a bicycle on a pump track. {M}aster thesis,
  {TU} {D}ortmund {U}niversity, 2021.

\bibitem{fliessbach_lehrbuch_theroetische_physik}
T.~Flie{\ss}bach.
\newblock {\em Lehrbuch zur theoretischen Physik}.
\newblock Springer Spektrum, BerlinHeidelberg, 2009.

\bibitem{bartelmann_theroetische_physik}
M.~Bartelmann, B.~Feuerbacher, T.~Kr{\"u}ger, D.~L{\"u}st, A.~Rebhan, and
  A.~Wipf.
\newblock {\em Theoretische Physik}.
\newblock Springer Berlin Heidelberg, 2015.

\bibitem{particle_surface}
T.~M{\"u}ller and J.~Frauendiener.
\newblock Charged particles constrained to a curved surface.
\newblock {\em European Journal of Physics}, 34(1):147--160, 1 2013.

\bibitem{wiki_velodrome}
Wikipedia.
\newblock {Velodrome} --- {Wikipedia}{,} the free encyclopedia.
\newblock https://en.wikipedia.org/wiki/Velodrome, 2022.
\newblock Accessed: 17.10.2023.

\bibitem{matlab}
MATLAB.
\newblock {\em version 9.10.0.1649659 (R2021a) Update 1}.
\newblock The MathWorks Inc., Natick, Massachusetts, 2021.

\bibitem{casad}
J.~A.~E. Andersson, J.~Gillis, G.~Horn, J.~B. Rawlings, and M.~Diehl.
\newblock {CasADi} -- {A} software framework for nonlinear optimization and
  optimal control.
\newblock {\em Mathematical Programming Computation}, In Press, 2018.

\bibitem{ipopt}
A.~W{\"a}chter and L.~T. Biegler.
\newblock On the implementation of an interior-point filter line-search
  algorithm for large-scale nonlinear programming.
\newblock {\em Mathematical programming}, 106(1):25--57, 2006.

\end{thebibliography}

\end{document}